\documentstyle[aps,twocolumn,epsfig]{revtex}
\begin{document}
\newcommand{\ve}[1]{\mbox{\boldmath $#1$}}
\twocolumn[\hsize\textwidth\columnwidth\hsize
\csname@twocolumnfalse%
\endcsname
 
\draft

\title {Weakly-interacting Bose-Einstein condensates under rotation} 
\author{G. M. Kavoulakis, B. Mottelson, and C. J. Pethick}
\date{\today} 
\address{NORDITA, Blegdamsvej 17, DK-2100 Copenhagen \O, Denmark}
\maketitle
 
\begin{abstract}

   We investigate the ground and low excited states of a rotating, 
weakly interacting Bose-Einstein condensed gas in a harmonic trap
for a given angular momentum. Analytical results in various limits,
as well as numerical results are presented, and these are compared
with those of previous studies.
Within the mean-field approximation and for repulsive
interaction between the atoms, we find that for very low values of
the total angular momentum per particle, $L/N \rightarrow 0$, where
$L \hbar$ is the angular momentum and $N$ is the
total number of particles, the angular momentum is carried by 
quadrupolar ($|m| = 2$) surface modes. For
$L/N =1$ a vortex-like state is formed and all the atoms occupy the $m=1$
state. For small negative values of $L/N-1$ the states with $m=0$ and $m=2$
become populated, and for small positive values of $L/N-1$ atoms in the states 
with $m=5$ and $m=6$ carry the additional angular momentum. In the whole 
region $0 \le L/N \le 1$ we have strong analytic and numerical evidence 
that the interaction energy drops linearly as a function of $L/N$. We have also
found that an array of singly quantized vortices is formed as $L/N$ increases. 
Finally we have gone beyond the mean-field approximation and have
calculated the energy of the lowest state up to order $N$ for small negative 
values of $L/N-1$, as well as the energy of the low-lying excited states. 

\end{abstract}
 
\vskip0.5pc]

\section{Introduction}

   One of the basic questions about Bose-Einstein condensates in
trapped alkali atom vapors \cite{RMP} is how they behave under rotation.
A lot of theoretical work has been done on this subject, both analytical, 
and numerical \cite{Rokhsarv,Wilkin,Rokhsar,Fetter,Ben,Bertsch,Feder}, and the
problem has been studied theoretically, both in the Thomas-Fermi limit of 
strong interactions \cite{Rokhsarv,Fetter,Feder} and in the limit of
weak interactions \cite{Wilkin,Rokhsar,Ben,Bertsch}, which we consider in 
this paper.

   In Ref.\,\cite{Wilkin} Wilkin {\it et al.} considered a weakly interacting
Bose gas with attractive interactions and showed that in the lowest
energy state of a given angular momentum,
the angular momentum is carried by the center of mass
motion. Butts and Rokhsar calculated numerically the moment of inertia
of a weakly-interacting trapped Bose gas with effective repulsive 
interactions \cite{Rokhsar}. One of us identified the elementary modes
of excitation for small angular momentum and demonstrated in Ref.\,\cite{Ben}
that a system of rotating weakly-interacting bosons exhibits two
additional kinds of condensation associated with the nature of low-lying
excitations. Finally Bertsch and Papenbrock performed in Ref.\,\cite{Bertsch}
exact numerical diagonalization
within the subspace of states with a given angular momentum, which are
degenerate in the absence of interactions.

   Experimentally the detection of vortex states
in a two-component system has been reported by Matthews {\it et al.}
\cite{JILA}, while Madison {\it et al.} \cite{Madison} have provided
evidence for the formation of vortex states in a stirred one-component
Bose-Einstein condensate.

   Our basic goal in this study is to identify the lowest energy
states of a harmonically trapped, weakly interacting Bose gas for
a given angular momentum $L$. As we show below these states are
selected by the interactions. In Sec.\,II we describe the model 
and discuss the degeneracy of the many-body states for
a given angular momentum in the absence of interactions. In Sec.\,III
we use the mean-field approximation to calculate the interaction energy, and 
derive numerical and analytical results under various conditions. In
Sec.\,IV we describe how one can go beyond the mean-field approximation
and study as an example the specific case of small negative $L/N-1$.
Finally in Sec.\,V we give our conclusions.

\section{The model}

  Our starting point is the Hamiltonian $H$, given by
\begin{eqnarray}
  H = H_0 + V.
\label{ham}
\end{eqnarray}
Here
\begin{eqnarray}
    H_0 = \sum_{i} - \frac {\hbar^{2}} {2M} {\ve \nabla}_{i}^{2} +
  \sum_i  \frac 1 2 \, M \omega^{2} (x_{i}^{2} + y_{i}^{2}) + f(z_i)
\label{h0}
\end{eqnarray}
includes the kinetic energy of the particles and their potential energy
due to the trapping potential. The axis of rotation is taken to be the
$z$ axis, and the trapping potential is assumed to be that of an
isotropic harmonic oscillator of frequency $\omega$ in the $x$-$y$ plane.  
Also $M$ is the mass of the atoms. Our results do not depend on the 
trapping potential $f(z)$ in the $z$ direction. The interaction $V$
between atoms is assumed to be of zero range,
\begin{eqnarray}
   V = \frac 1 2 U_{0} \sum_{i \neq j} \delta({\bf r}_{i} - {\bf r}_{j}),
\label{v}
\end{eqnarray}
where $U_0 = 4 \pi \hbar^2 a/M$ is the strength 
of the effective two-body interaction, with $a$ being the scattering 
length for atom-atom collisions. We assume that the interaction is repulsive,
$a > 0$. Attractive interactions have been studied in 
Refs.\,{\cite{Wilkin,Ben}.

Much theoretical work on rotating condensates has been done in the 
Thomas-Fermi limit of strong interactions, where the superfluid coherence 
length
\begin{eqnarray}
    \xi = (8 \pi n a)^{-1/2},  
\label{ksi}
\end{eqnarray}
$n$ being the particle density, is much less than the size
of the cloud. Under these conditions the system is expected to
exhibit superfluid properties much like those of liquid helium II \cite{GC}. 
In this study we examine the opposite limit of weak interactions,
$n U_0 \ll \hbar \omega$ and $n U_0 \ll \Delta E_z$, where 
$\Delta E_z$ is the energy separation between the first excited
state and the ground state for motion in the $z$ direction. Under the 
above conditions
\begin{eqnarray}
   \frac {\xi} {a_{\rm osc}} \sim  \left(
  \frac {a_z} {N a} \right)^{1/2} ,
\label{cond}
\end{eqnarray}
where $N$ is the number of atoms in the trap, $a_{\rm osc} = 
(\hbar/M \omega )^{1/2}$ is the oscillator length, and 
$a_z$ is the characteristic length associated with the 
motion of the atoms along the $z$ axis. Therefore the coherence
length is larger than the size of the cloud,
and the situation is analogous to that for BCS pairing of nucleons in nuclei.

   Since we consider rotation around the $z$ axis, the condition 
$n U_0 \ll \Delta E_z$ implies that the motion along this axis is
frozen out and the problem is essentially two-dimensional.
It is well known that for the harmonic oscillator potential in two 
dimensions the single-particle energies $\epsilon$ are given in the absence of
interactions by
\begin{eqnarray}
  \epsilon = (2 n_r + |m| + 1) \hbar \omega,
\label{energy}
\end{eqnarray}
where $n_r$ is the radial quantum number, and $m$ is the quantum number
corresponding to the angular momentum. In the lowest energy state of
the many boson system all particles are in states with $n_r = 0$,
and with $m$ being zero or having the same sign as the total angular 
momentum. The energy of the lowest state
of a system of non-interacting bosons with angular
momentum $L$ measured relative to that of the ground state is 
therefore
\begin{eqnarray}
    E = L \hbar \omega.
\label{energytni}
\end{eqnarray}
There is a huge degeneracy 
corresponding to the many different ways of distributing $L$ quanta
of angular momentum among $N$ atoms.
Interactions between the atoms lift the degeneracy. We incorporate
the effect of the interactions in both the mean-field approximation, 
as well as by diagonalization within some appropriately chosen 
truncated space of degenerate states. We describe the two methods
separately below.
\noindent
\begin{figure}
\begin{center}
\epsfig{file=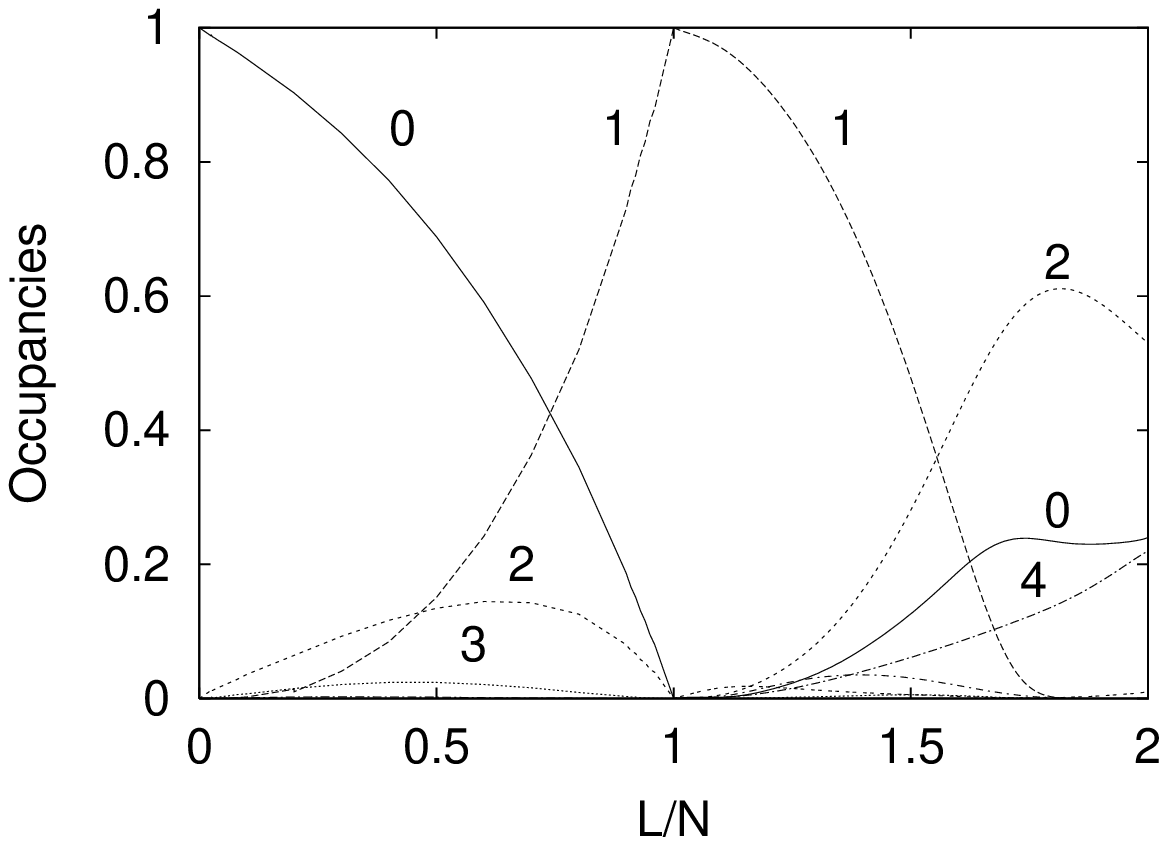,width=6.0cm,height=6.0cm,angle=0}
\begin{caption}
{The numerical result for $|c_m|^2$ which comes from the minimization of the
energy, as a function of $L/N$. The numbers refer to the corresponding states
with angular momentum $m \hbar$. The lowest two curves in the region $L/N 
\approx 1.5$ give the occupancy of the $m=5$ (higher) and $m=6$ (lower) 
states.} 
\end{caption}
\end{center}
\label{FIG1}
\end{figure}
\begin{figure}
\begin{center}
\epsfig{file=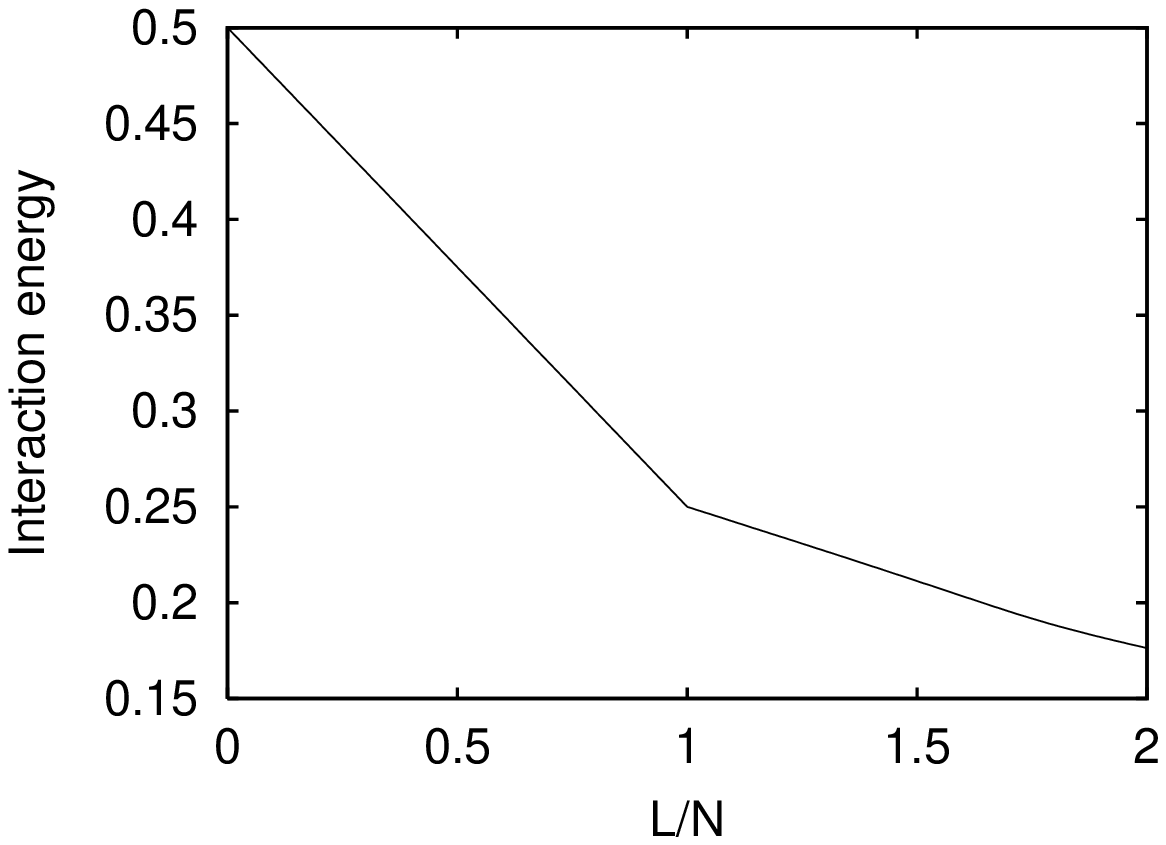,width=6.0cm,height=6.0cm,angle=0}
\begin{caption}
{The interaction energy, $\langle V \rangle$, in units of $N^2 v_0$
as a function of $L/N$.}
\end{caption}
\end{center}
\label{FIG2}
\end{figure}
\noindent

\section{Mean-field approximation}

   We start with the mean-field Gross-Pitaevskii approach. Butts and
Rokhsar have used this method to derive numerical results for the moment
of inertia of a Bose gas \cite{Rokhsar}.
In this scheme the many-body condensate wavefunction with $N$ particles
and $L$ units of angular momentum
$\Psi_{L,N}({\bf r}_1,{\bf r}_2,\ldots,{\bf r}_N)$
is taken to be the product of the single-particle states $\Psi({\bf r}_i)$,
\begin{eqnarray}
    \Psi_{L,N}({\bf r}_1,{\bf r}_2,\ldots,{\bf r}_N) = 
  \Psi({\bf r}_1) \times \Psi({\bf r}_2) \ldots \Psi({\bf r}_N).
\label{nwf}
\end{eqnarray}
The single-particle states $\Psi({\bf r}_i)$ can be expanded in terms
of the harmonic-oscillator eigenstates $\Phi_{m}({\bf r}_i)$: 
\begin{eqnarray}
   \Psi({\bf r}_i) = \sum_{m=0}^{\infty} c_{m} \Phi_{m}({\bf r}_i),
\label{exp}
\end{eqnarray}
where the $c_{m}$ are variational parameters, which are complex in general
and are functions of $L$. The summation in Eq.\,(\ref{exp}) is restricted 
to positive $m$, since states with negative $m$ do not belong to 
the space of degenerate states. The quantity $|c_{m}|^2$ gives the
occupation probability for state $\Phi_m$. Also
\begin{equation}
    \Phi_m({\bf r}) = \frac 1 {(m! \pi a_{\rm osc}^2)^{1/2}} \, g(z)
   \left( \frac {\rho}{a_{\rm osc}} \right)^{|m|} e^{i m \phi} 
 e^{-\rho^{2}/2 a_{\rm osc}^2}. 
\label{phim}
\end{equation}
Here $\rho, z$, and $\phi$ are cylindrical polar coordinates. In the above
expression we have assumed that the bosons are in their ground state
$g(z)$ along the axis of rotation. The expectation value 
of the interaction energy $V$ in the state given by Eq.\,(\ref{nwf}) is
\begin{eqnarray}
   \langle V \rangle = \frac 1 2 N (N-1) U_{0}
\int |\Psi|^{4} \, d{\bf r}.
\label{e}
\end{eqnarray}
To find the lowest energy state we calculate $\langle V \rangle$
as a function of the variational parameters $c_{m}$, and minimize it with 
respect to them under the following two constraints: the normalization
condition,
\begin{eqnarray}
  \sum_{m} |c_{m}^{2}| = 1,
\label{cond1}
\end{eqnarray}
and the condition that the expectation value of the angular momentum per 
particle be fixed,
\begin{eqnarray}
  \sum_{m} m |c_{m}|^{2} = L/N.
\label{cond2}
\end{eqnarray}
The parameters $c_m$ are complex in general, and therefore both their
magnitudes, and their phases need to be determined. However 
Eqs.\,(\ref{cond1}) and (\ref{cond2}) impose two constraints on the magnitudes
of the $c_m$. Furthermore, the overall phase of the wavefunction is
arbitrary. Finally the rotational symmetry of the confining potential implies
that the origin of the angular coordinate is arbitrary, which corresponds
to the condition for conservation of angular momentum, which holds
even in the presence of interactions.
Therefore if the expansion (\ref{exp}) is truncated at a value
$m_{\rm max}$, the number of independent variables is $2 \times
(m_{\rm max}+1) - 4 = 2(m_{\rm max}-1)$.

\subsection{Numerical results}

   We have examined the problem numerically with $m_{\rm max}$ up to 9.
The total number of terms in the expression for $\langle V \rangle$
is 125 in this case. The result of such a calculation with
$m_{\rm max} = 6$ is shown in Fig.\,1 for $0 \le L/N \le 2$. 
\noindent
\begin{figure}
\begin{center}
\epsfig{file=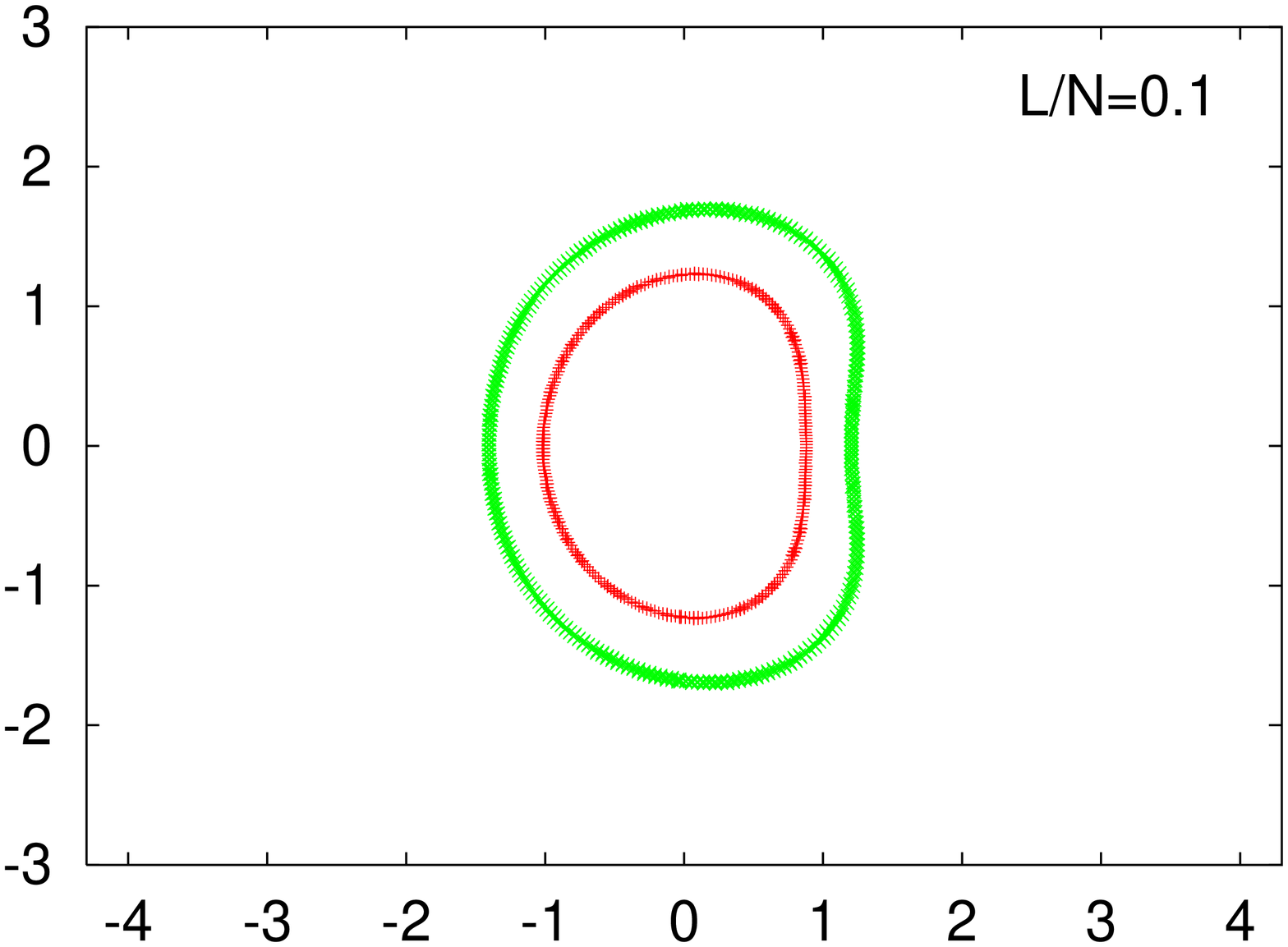,width=3.0cm,height=4.2cm,angle=-90}
\epsfig{file=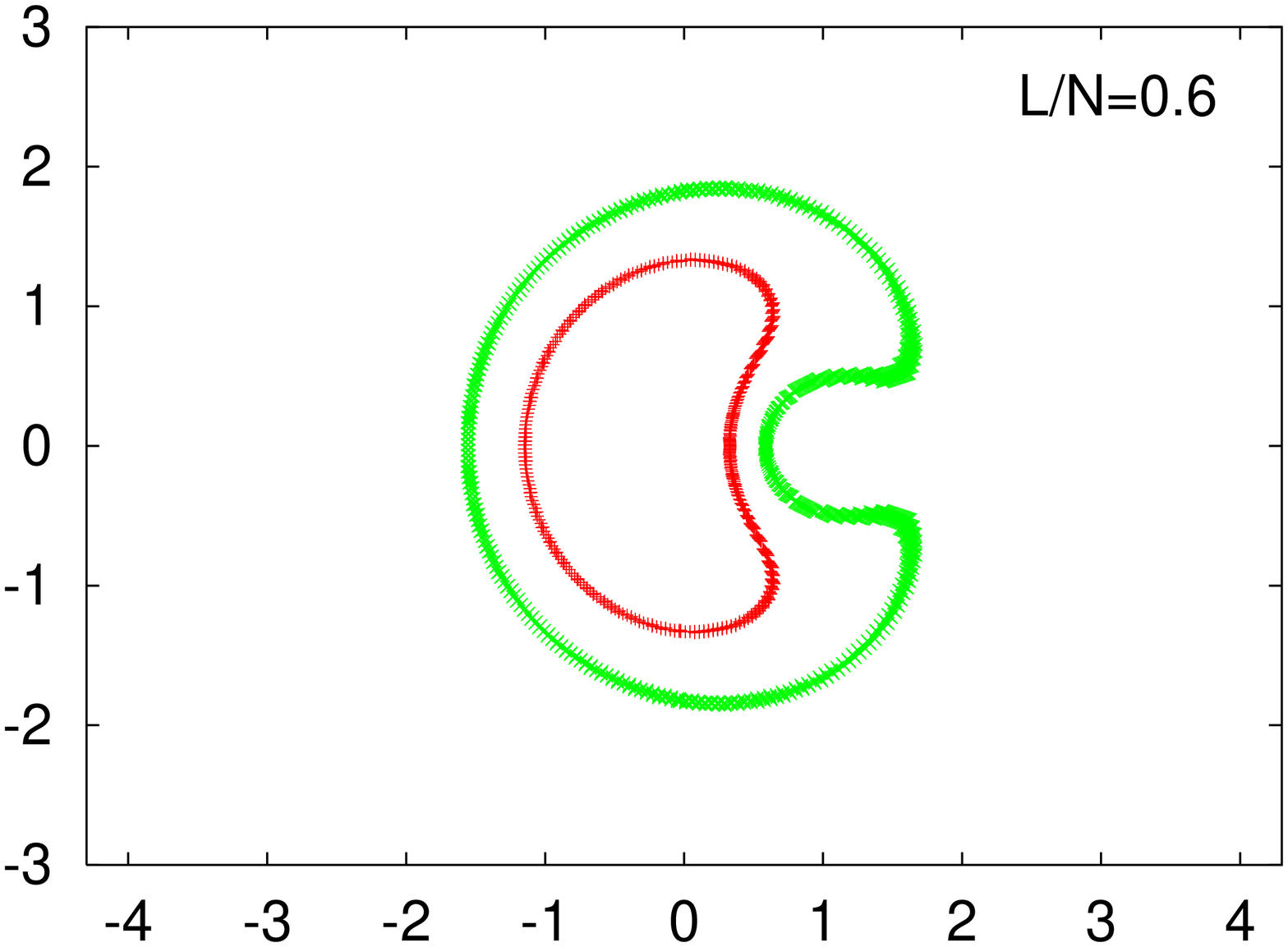,width=3.0cm,height=4.2cm,angle=-90}
\epsfig{file=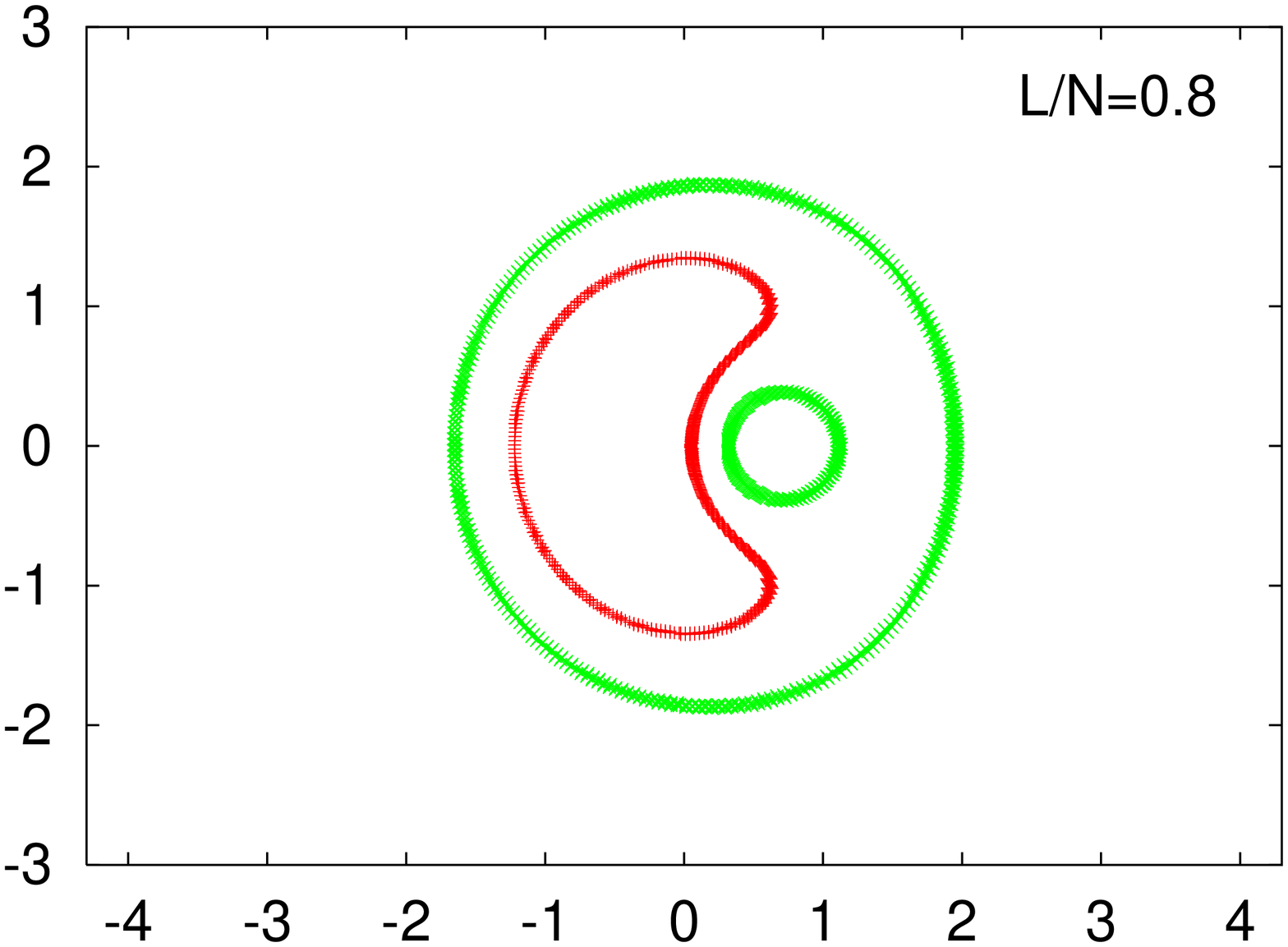,width=3.0cm,height=4.2cm,angle=-90}
\epsfig{file=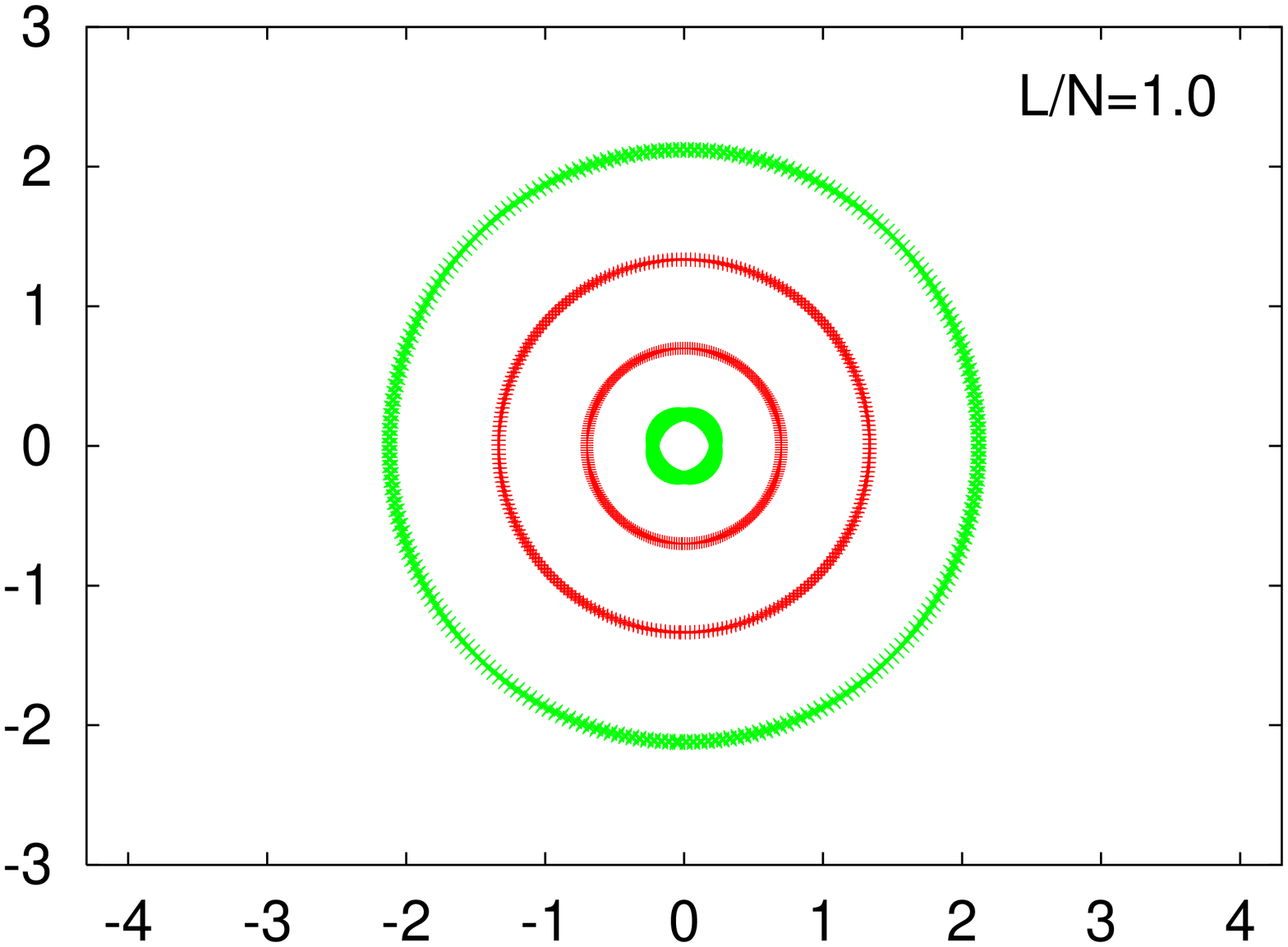,width=3.0cm,height=4.2cm,angle=-90}
\vskip 0.5pc
\begin{caption}
{Lines of constant density for $L/N = 0.1, 0.6, 0.8,$ and 1.0 in
a plane perpendicular to the $z$ axis. On the darker curves
$|\Psi|^2=0.3 n_0$, and on the lighter curves $|\Psi|^2=0.1 n_0$,
where $n_0 = g(z)^2/\pi a_{\rm osc}^2$. 
The unit of length is the oscillator length $a_{\rm osc}$. These 
pictures show how a vortex enters the cloud of bosons
as the angular momentum per particle increases.}
\end{caption}
\end{center}
\label{FIG3}
\end{figure}
\noindent
We show the results with $m_{\rm max} = 6$, since the occupancy
of states with higher $m$ is very low, and therefore including
such states would not alter the results on this scale.
Figure 2 shows the corresponding interaction energy.
Also Fig.\,3 shows the lines of constant density, $|\Psi|^2=$ constant
for $L/N = 0.1, 0.6, 0.8,$ and 1.0. Figure 3 shows the gradual transition 
from mostly quadrupole and to a less extent octupole excitations,
which are present at low angular momentum, to vortex-like
structures as $L$ approaches $N$.
We should also mention that the structures in Fig.\,3, as well as
those in Figs.\,4, 5 and 6, rotate with an angular frequency $\Omega$ given by
\begin{eqnarray}
   \Omega = \frac 1 {\hbar} \, \frac {\partial E_{\rm tot}} {\partial L}
 = \omega - \frac 1 {N \hbar} \,
 \frac {\partial \langle V \rangle} {\partial l},
\label{omega}
\end{eqnarray}
which is lower than the trap frequency $\omega$. Here $E_{\rm tot}$
is the total energy of the system.

  When $L$ increases beyond $N$, the rotational invariance for $L/N=1$
is lost. Density contours for various values of $L$ between $N$ and
$2N$ are shown in Fig.\,4. These were calculated including the states
up to $m=6$. For $L \agt 1.75 N$ the optimal wavefunction has a two-fold
axis of symmetry, and the odd-$m$ coefficients in the wavefunction
vanish smoothly as the transition is approached, as shown in Fig.\,1.
In Figs.\,5 and 6 we show contours for $L=2N$ and $L=2.1N$. There is a 
first-order transition from a state with two-fold symmetry to one with 
three-fold symmetry for $L \approx 2.03 N$. The solution for $L/N=2.0$ 
with the states $m=0$, 2, 4, 6, and 8 considered has an energy of
$\approx 0.1757 N^2 v_0$, whereas the one with the three-fold symmetry,
with $m=0$, 3, 6, and 9, has $0.1761 N^2 v_0$. In contrast for $L/N = 2.1$,
the state with the three-fold symmetry has an energy $0.1691 N^2 v_0$,
\begin{figure}
\begin{center}
\epsfig{file=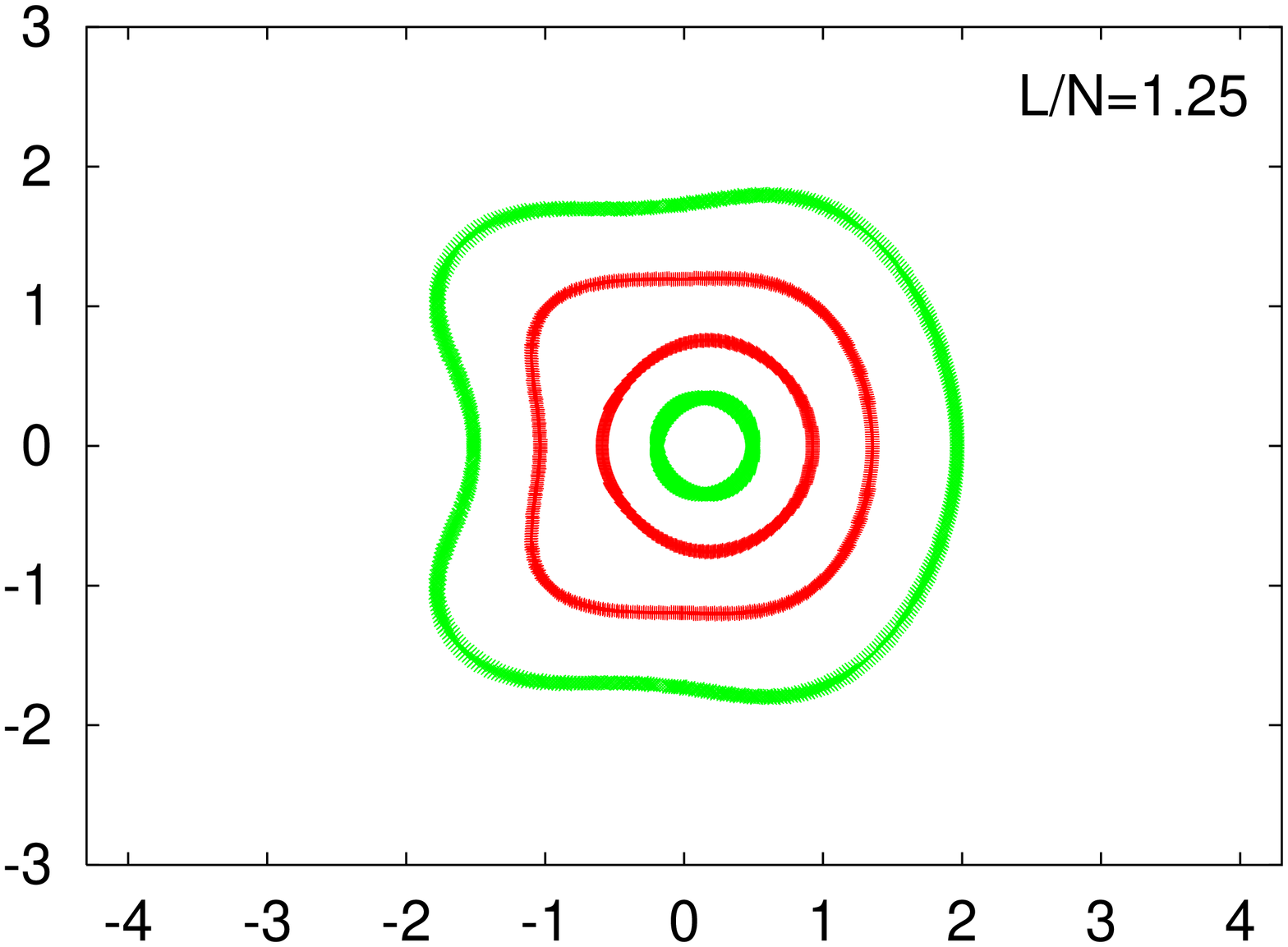,width=3.0cm,height=4.2cm,angle=-90}
\epsfig{file=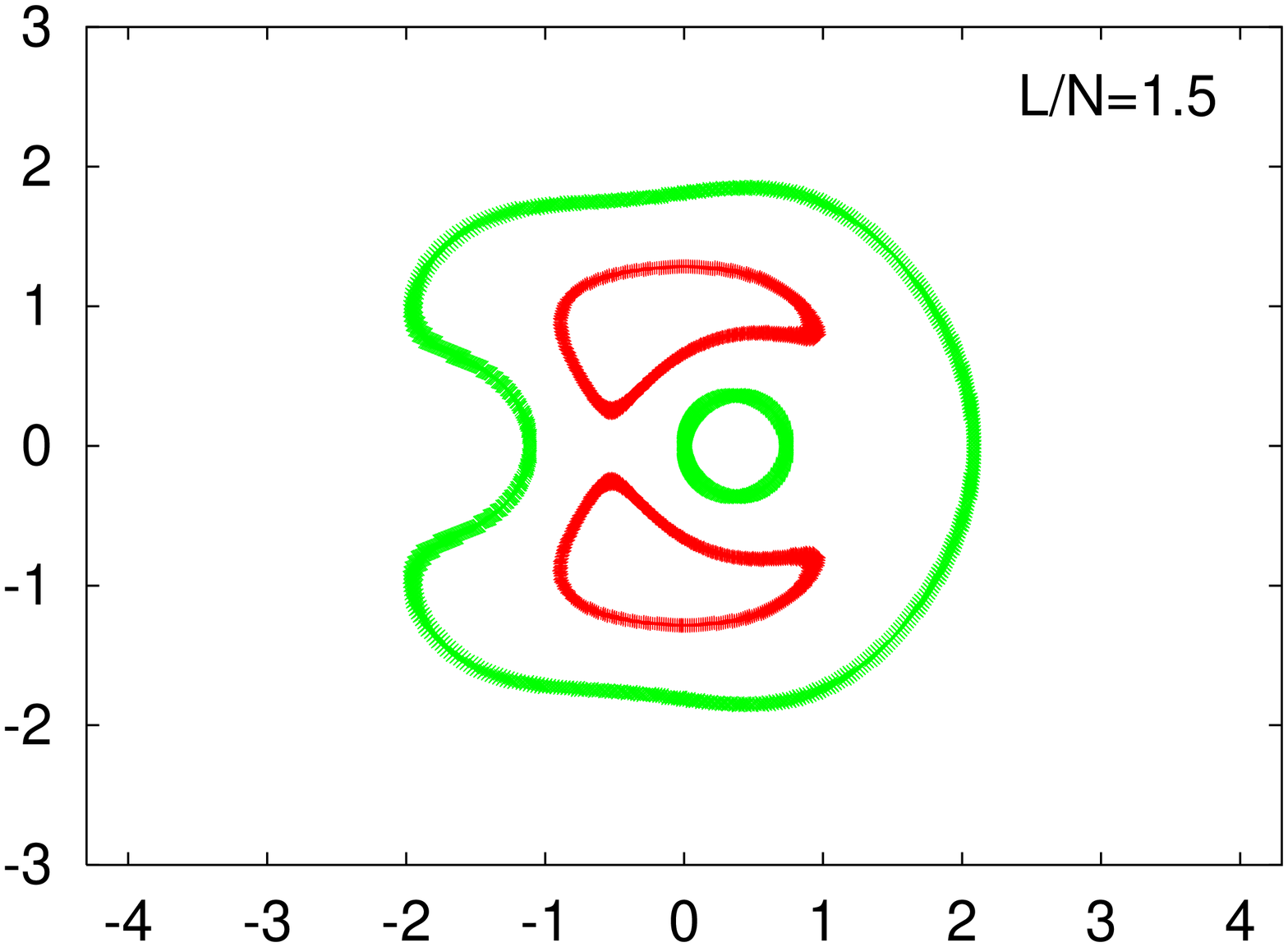,width=3.0cm,height=4.2cm,angle=-90}
\epsfig{file=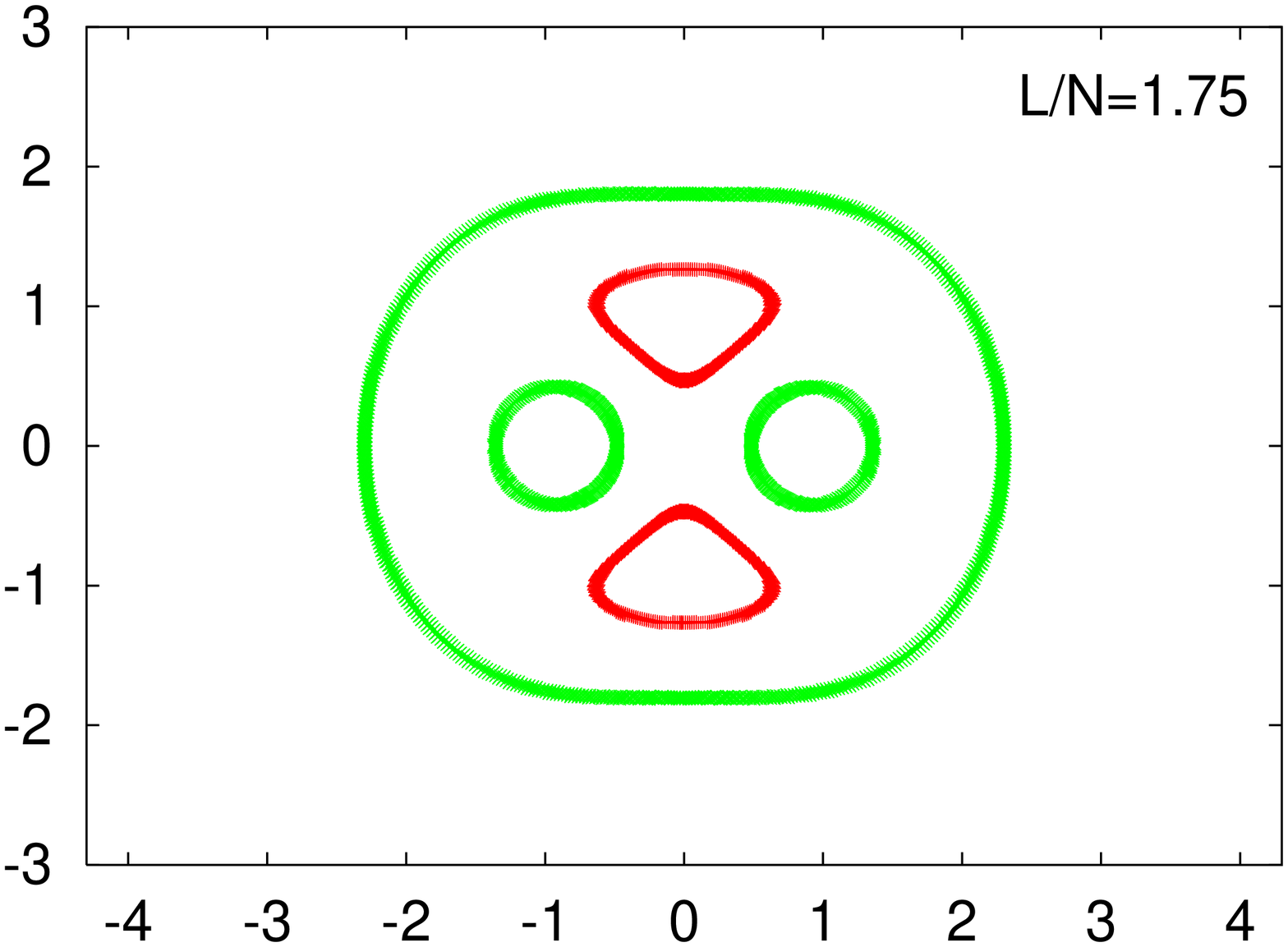,width=3.0cm,height=4.2cm,angle=-90}
\epsfig{file=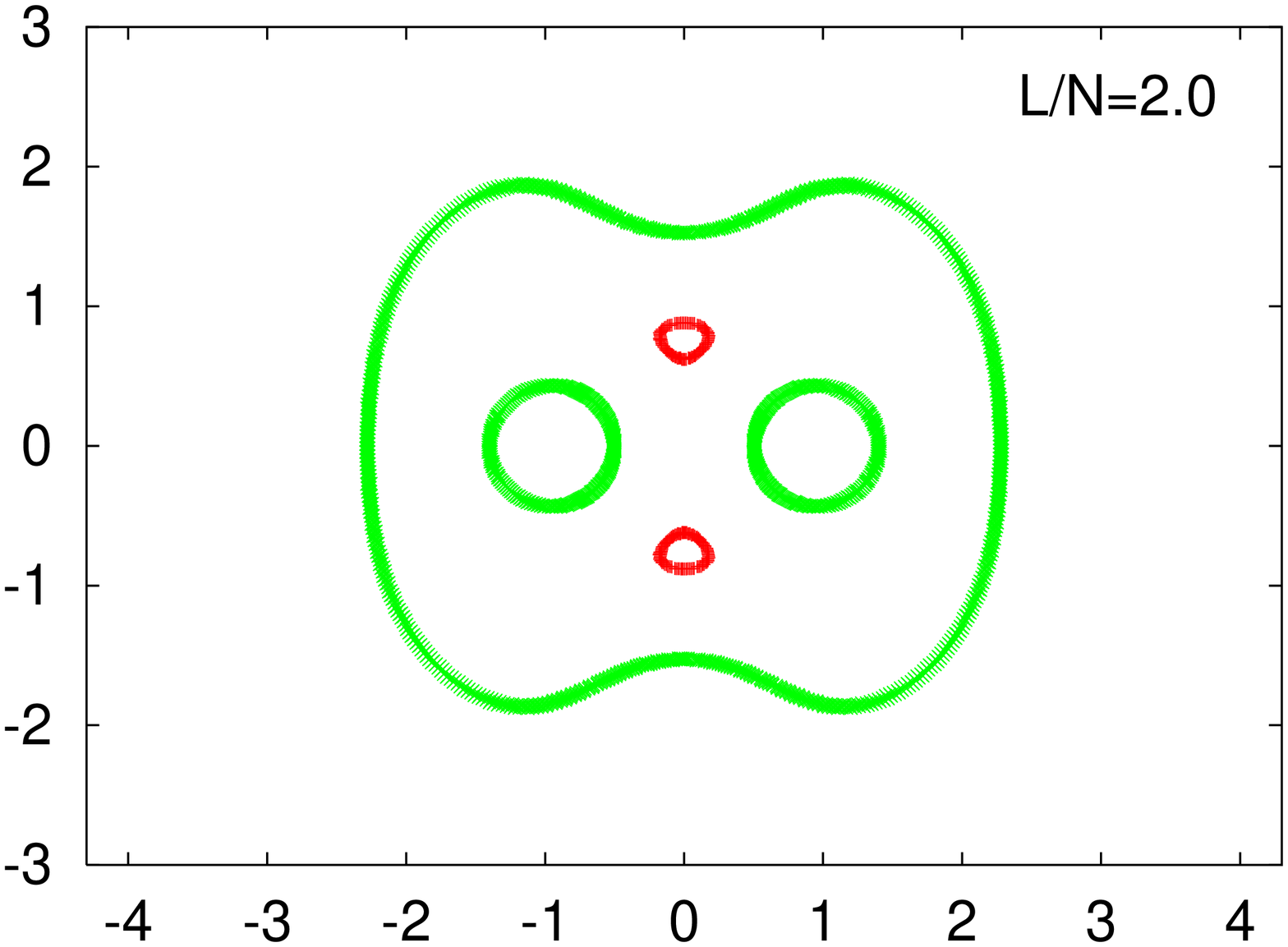,width=3.0cm,height=4.2cm,angle=-90}
\epsfig{file=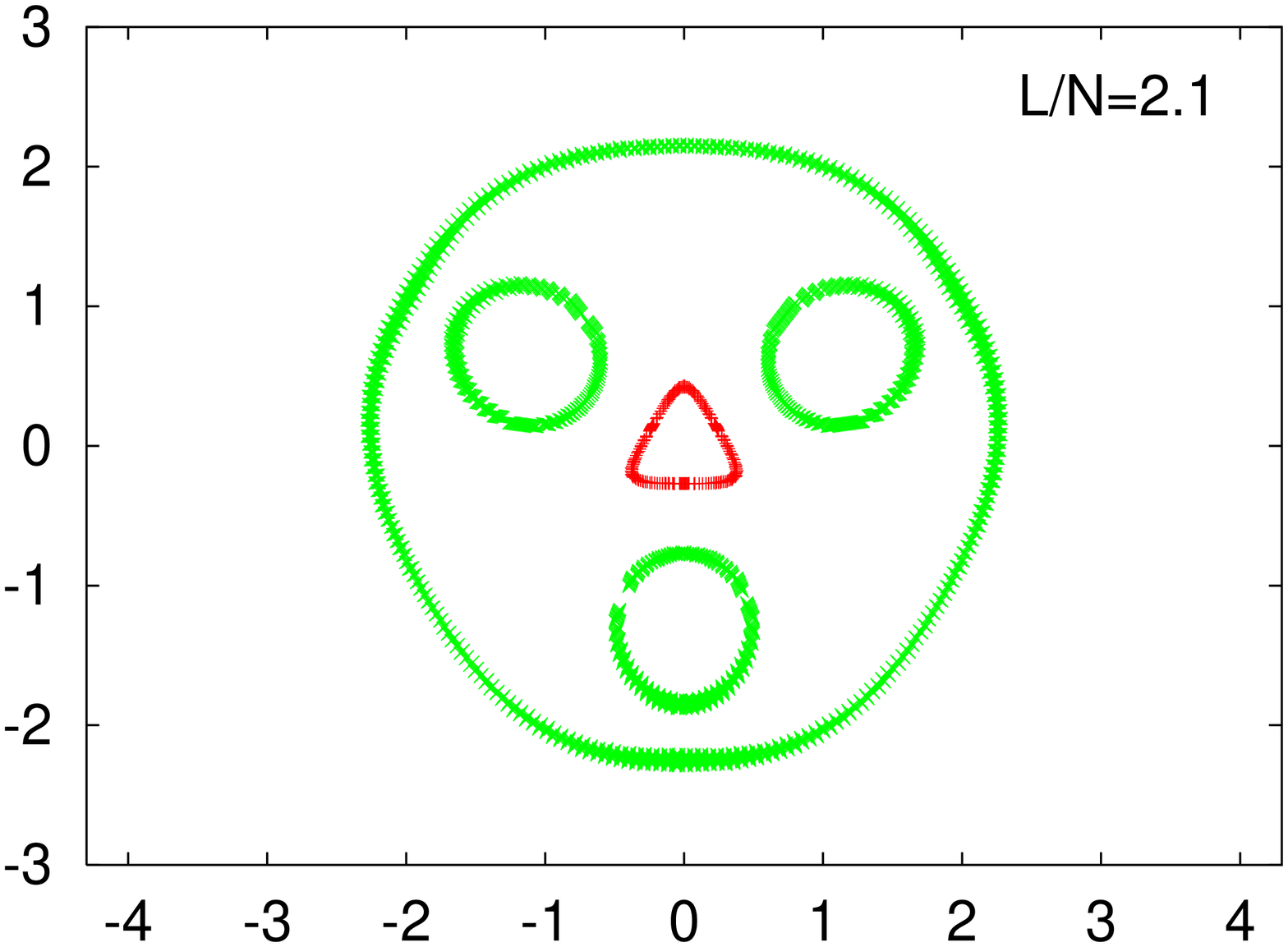,width=3.0cm,height=4.2cm,angle=-90}
\vskip 0.5pc
\begin{caption}
{Lines of constant density for $L/N = 1.25, 1.5$, 1.75 2.0, and 2.1
in a plane perpendicular to the $z$ axis. On the darker curves
$|\Psi|^2=0.3 n_0$, and on the lighter curves $|\Psi|^2=0.1 n_0$.}
\end{caption}
\end{center}
\label{FIG4}
\end{figure}
\noindent
which is lower than the solution with the two-fold symmetry, with an energy
$0.1700 N^2 v_0$.

More generally, we have found that as $L/N$ increases, 
the lowest-energy states are the ones where a vortex array is formed,
in agreement with the results of Ref.\cite{Rokhsar}. 

\subsection{Analytical results for $L/N \rightarrow$ 0}

   We now turn to an analytic approach to the problem.
One can systematically develop a power-series expansion for the occupancies
$|c_m|^2$ of the states, as well as for the energy in certain limits.
We start with the case of very low angular momentum, $l = L/N \rightarrow 0$.
Working in terms of quantum-mechanical states, one of us has calculated
in Ref.\,\cite{Ben} the difference in the energies
of two states where in the one state all $N$ particles have $m=0$ and
in the other state a particle is promoted to the state with
$m=\lambda$, and $N-1$ particles have $m=0$. If we denote the states as
\begin{eqnarray}
  |0^{N_0}, 1^{N_1}, 2^{N_2}, \dots \rangle, 
\label{states}
\end{eqnarray}
where $N_m$ is the number of particles with angular momentum $m \hbar$,
the two states are
$  |0^{N}\rangle,$ and $|0^{N-1}, \lambda^{1} \rangle$.
The difference $\epsilon_{\lambda}$ in the energy  
between these two states corresponding to this $2^{\lambda}$-pole excitation 
is given by
\begin{eqnarray}
    \epsilon_{\lambda} =  \lambda \hbar \omega
   - \left( 1 - \frac 1 {2^{\lambda -1}} \right) N v_{0} + {\cal O} (v_0),
\label{exce}
\end{eqnarray}
where $v_0 = U_0 \int |\Phi_{0}|^{4} \, d{\bf r}$. One can easily see
from Eq.\,(\ref{exce}) that at this level of approximation
the excitations with the highest gain in 
interaction energy per unit of angular momentum are the ones with
$\lambda = 2$ or $\lambda = 3$, i.e., quadrupole or octupole excitations.

  We now calculate the interaction energy for low values of $l$. The 
calculation of the energies of elementary excitations indicates that one 
would expect quadrupolar and octupolar modes to be the most important
ones for small $l$. To determine the most energetically favorable
way of giving the system angular momentum, one has to identify
the behavior of $|c_2|^2$ and $|c_3|^2$ as $l = L/N \rightarrow 0$, and then
it is possible to build up a whole power-series expansion. Motivated by 
the fact that the $\lambda=2$ and 3 excitations are degenerate,
and are the ones which give the highest gain in energy per unit of 
angular momentum, we assume that both $c_2$ and $c_3$ are of order 
$l^{1/2}$. As we show below, it is the mode-mode interaction that 
lifts this degeneracy, making the $\lambda=2$ mode dominant for low 
values of angular momentum. 

It is instructive to
give an explicit example, so let us assume that we wish to examine the 
interaction energy up to order $l^2$. In order to minimize the interaction 
energy to this order, the states with $m=1$, 4, 5, and 6 need to be
considered, since the phases of off-diagonal terms, like for example
$|c_0| |c_1| |c_2| |c_3|$, can be chosen to have a negative sign,
and thus lower the energy as compared to the case where only 
$c_0$, $c_2$, and $c_3$ are non-zero. A useful formula for the 
matrix elements of the potential is 
\begin{eqnarray}
   \int \Phi_k^*({\bf r}) \Phi_l^*({\bf r}) \Phi_m({\bf r})
  \Phi_n({\bf r}) \, d{\bf r} =  \phantom{XXX}
\nonumber \\ \phantom{XXX}
  \delta_{k+l,m+n} \, \frac {(k+l)!} {2^{(k+l)}
 \sqrt{k!\, l! \, m! \, n!}} \int |\Phi_0({\bf r})|^4 d{\bf r}.
\label{integral}
\end{eqnarray}
As will become clear below, to calculate the energy up to order $l^2$ 
we must include the following terms:
\begin{eqnarray}
    \langle V \rangle = \left( \frac 1 2 |c_0|^4 + \frac 1 2 |c_0|^2 |c_2|^2
   + \frac 1 4 |c_0|^2 |c_3|^2  \right.
\nonumber \\
+ \frac 3 {16} |c_2|^4 
  + \frac 5 {32} |c_3|^4  +  \frac 5 8 |c_2|^2 |c_3|^2
\nonumber \\
 + |c_0|^2 |c_1|^2  -  \frac {\sqrt 3} 2 |c_0| |c_1| |c_2| |c_3|
\phantom{XXXXX}
\nonumber \\
 + \frac 1 8 |c_0|^2 |c_4|^2  -  \frac {\sqrt 6} 8 |c_0| |c_2|^2 |c_4|
\phantom{XXXXXX}
\nonumber \\
 + \frac 1 {16} |c_0|^2 |c_5|^2 - \frac {\sqrt {10}} 8 |c_0| |c_2| |c_3| |c_5|
\phantom{XXXXXX}
\nonumber \\ \left.
 + \frac 1 {32} |c_0|^2 |c_6|^2  -  \frac {\sqrt 5} {16} |c_0| |c_3|^2 |c_6|
 \right) N^2 v_0 + {\cal O}(N v_0).
\label{lsqr}
\end{eqnarray}  
In the above expression we have chosen the phases $\phi_m$ of the 
variational coefficients $c_m$ in such a way as to minimize
$\langle V \rangle$, and in the specific example we can
arrange them so that all the off-diagonal matrix elements are
negative. One of the phases can have any value, and we 
make the choice $\phi_0=0$. The rest of them can be expressed
in terms of, say, $\phi_1$. We have found that up to $m=6$ [$m \neq 0]$
the expression
\begin{eqnarray}
  \phi_m = m \phi_1 + (m+1) \pi
\label{phases}
\end{eqnarray}
gives the lowest energy.

It is convenient to
introduce the variable $X = |c_2|^2 + |c_3|^2$, which is linear in $l$ 
to leading order, and make use
of the constraints given by Eqs.\,(\ref{cond1}) and (\ref{cond2}) to get  
\begin{eqnarray}
  |c_0|^2&=&1-X - |c_1|^2 - |c_4|^2 - |c_5|^2 - |c_6|^2; 
\nonumber \\ 
  |c_2|^2&=&3X-l + |c_1|^2 + 4 |c_4|^2 + 5 |c_5|^2 + 6 |c_6|^2; 
\nonumber \\
 |c_3|^2&=&l-2X - |c_1|^2 - 4 |c_4|^2 - 5 |c_5|^2 - 6 |c_6|^2.
\label{c0c2c3}
\end{eqnarray}
Then Eq.\,(\ref{lsqr}) takes the form 
\begin{eqnarray}
    \langle V \rangle = \left( \frac 1 2 - \frac l 4 
   -\frac {31} {16} X^2  + \frac {13} {8} l X - \frac {9} {32} l^2 \right.
\nonumber \\
 + \frac 1 4 |c_1|^2  -  \frac {\sqrt 3} 2 |c_1| |c_2| |c_3|
\phantom{XXXXX}
\nonumber \\
 + \frac 1 8 |c_4|^2  -  \frac {\sqrt 6} 8 |c_2|^2 |c_4|
\phantom{XXXXXX}
\nonumber \\
 + \frac 5 {16} |c_5|^2 - \frac {\sqrt {10}} 8 |c_2| |c_3| |c_5|
\phantom{XXXXXX}
\nonumber \\ \left.
 + \frac {17} {32} |c_6|^2  -  \frac {\sqrt 5} {16} |c_3|^2 |c_6|
 \right) N^2 v_0 + {\cal O}(N v_0).
\label{lsqrr}
\end{eqnarray}
The last four terms in the above equation can lower the energy
to order $l^2$. For $c_4$, for example,
the energy is minimized if [see the third line of Eq.\,(\ref{lsqrr})]
\begin{eqnarray}
  \frac {\partial} {\partial |c_4|} \frac 1 8 |c_4|^2  
 = \frac {\partial} {\partial |c_4|} \frac {\sqrt 6} 8 |c_2|^2 
 |c_4|,
\label{c1enec}
\end{eqnarray}
or
\begin{eqnarray}
   |c_4| = \frac {\sqrt 6} 2 |c_2|^2 \propto l.
\label{c1enecc}
\end{eqnarray}
Due to the non zero value of $c_4$ the energy is lowered by an amount
\begin{eqnarray}
   \Delta {\cal E} &=& \left( \frac 1 8 |c_4|^2  -  
  \frac {\sqrt 6} 8 |c_2|^2 |c_4| \right) N^2 v_0
\nonumber \\
 &=& - \frac 3 {16} |c_2|^4 N^2 v_0 \propto l^2 N^2 v_0.
\label{corrc4}
\end{eqnarray}
It is remarkable that the term $\Delta {\cal E}$ exactly cancels the
term $3 |c_4|^4/16$ in the second line of Eq.\,(\ref{lsqr}). 
In a similar way $c_1$, $c_5$, and $c_6$ can be expressed
in terms of $c_2$ and $c_3$ (and thus $X$), and Eq.\,(\ref{lsqr}) 
takes the form of the effective Hamiltonian
\begin{eqnarray}
  \langle V \rangle = \left( \frac 1 2 |c_0|^4 + \frac 1 2 |c_0|^2 |c_2|^2
   + \frac 1 4 |c_0|^2 |c_3|^2 \right.
\nonumber \\
  \left. + \frac 5 {34} |c_3|^4  -  \frac 1 4 |c_2|^2 |c_3|^2 \right) N^2 v_0
 + {\cal O} (N v_0),
\label{lsqreff}
\end{eqnarray}
or equivalently
\begin{equation}
  \langle V \rangle = \left[ \frac 1 2 - \frac l 4 + \frac {27} {17} 
  \left( X-\frac l 2 \right)^2 \right] N^2 v_0
 + {\cal O} (N v_0).
\label{lsqrefff}
\end{equation}
Minimizing the above expression with respect to $X$
we find that $X = l/2$ and thus the angular 
momentum has to be carried by the $m=2$ state alone, since $|c_2|^2 = l/2$
and $|c_3|^2 = 0$ up to terms linear in $l$. Also the quadratic correction 
to $\langle V \rangle$ vanishes. Therefore for $L/N \rightarrow
0$, the quadrupole ($\lambda = 2$) excitations are dominant. This is
one of the important conclusions of our study.
We show in the Appendix that a diagrammatic perturbative expansion
which assumes that only the states with $m=0, 2$, and 3 are
occupied by a macroscopic number of particles, while all the other states
are not, [but still contribute to the energy] gives the same result.
If one goes to higher order in $l$ the interaction energy has within the
perturbative scheme a term of the form $|c_2|^3 |c_3|^2$, which includes
all the processes that convert three $\lambda = 2$ excitations to two
$\lambda = 3$ excitations. This term can combine with the term $|c_3|^4$,
which implies that it is possible that $|c_3|^2 \propto |c_2|^3 \propto 
l^{3/2}$, which actually turns out to be the case. Then $c_1$,
for example, is given according to the second line of Eq.\,(\ref{lsqrr}), by 
\begin{eqnarray}
   |c_1| =  \sqrt 3 |c_2| |c_3| \propto l^{5/4}.
\label{c1eneccc}
\end{eqnarray}
Using similar arguments we find that
\begin{eqnarray}
     |c_{m}|^{2} \propto l^{m/2} \, {\rm for} \, m \neq 1, \,{\rm and}  
  \, \, |c_{1}|^{2} \propto l^{5/2}.
\label{ress0}
\end{eqnarray}
The leading terms in $|c_m|^2$ are given by
\begin{eqnarray}
  |c_{0}|^{2}&=&1 - \frac 1 2 \, l+ \frac 1 3 \, l^{3/2},
 \nonumber \\
  |c_{1}|^{2}&=& l^{5/2} + 2 \, l^3,
  \nonumber \\
  |c_{2}|^{2}&=&\frac 1 2 \, l - l^{3/2},
  \nonumber \\
  |c_{3}|^{2}&=& \frac 2 3 \, l^{3/2},
  \nonumber \\
  |c_{4}|^{2}&=&\frac 3 8 \, l^{2} - \frac 3 2 \, l^{5/2} -
 \frac {1173} {816} \, l^{3},
  \nonumber \\
  |c_{5}|^{2}&=& \frac 2 {15} \, l^{5/2} - \frac 4 {15} \, l^3,
  \nonumber \\
 {\rm and \phantom{XX}}  |c_{6}|^{2}&=& \frac 1 {144} \, l^{3},
\label{cisl00}
\end{eqnarray}
and the corresponding interaction energy is
\begin{eqnarray}
  \langle V \rangle = \left[ \frac 1 2 - \frac l 4 + {\cal O} \, (l^4) \right]
 N^2  v_{0} + {\cal O} (N v_{0}).
\label{intel00}
\end{eqnarray}
The above equation is another basic result of our study, namely
that the interaction energy drops linearly with the angular momentum
up to the order we have examined, for $L/N \rightarrow 0$, in agreement
with our numerical simulations and with those of 
Refs.\,\cite{Rokhsar,Bertsch}.

\subsection{Analytical results for $l \approx 1$}

  We now turn to the region $L/N \approx 1$. When the angular momentum per 
particle is exactly equal to 1, the lowest-energy state is the one
where $|c_{1}|^{2} =1$, and corresponds to a vortex state.
We consider the two cases $l<1$ and $l>1$ separately, starting
with $l<1$. 
\vskip1pc
{\bf a. Analytical results for $l < 1$}
\vskip1pc
  The simplest way to create a state with $l < 1$ from that with $l=1$
is to transfer particles from the $m=1$ state to the ground state.
However, the energy can be even lower if also
the $m=2$ state is populated, because of the off-diagonal term
$|c_0| |c_1|^2 |c_2|$. The interaction energy up to order 
$\bar{l}= 1 - L/N$ is found by minimizing the potential energy,
retaining only the coefficients $c_0$, $c_1$, and $c_2$. This is
\begin{eqnarray}
   \langle V \rangle = \left( \frac 1 4 |c_1|^4 + |c_0|^2 |c_1|^2 +
  \frac 3 4 |c_1|^2 |c_2|^2 \right.
\nonumber \\ \left.
    - \frac {\sqrt 2} 2 |c_0| |c_1|^2 |c_2| \right) N^2 v_0
 + {\cal O} (N v_0), 
\label{enbuv}
\end{eqnarray}
where we have used the fact that for this case too the phases may be 
shown to be given by Eq.\,(\ref{phases}).
Equation (\ref{phases}) is valid for small negative $L/N - 1$  
at least up to $m=4$. Thus in this limit
\begin{eqnarray}
  |c_{0}|^{2} \propto |c_{2}|^{2} \propto \bar{l}.
\label{enbuvvv}
\end{eqnarray}
To obtain the coefficients of proportionality it is 
convenient to use the following parametrization:
\begin{eqnarray}
  |c_{0}|^{2}&=&(1+\alpha) \bar{l},
\nonumber \\
  |c_{1}|^{2}&=&1 - (1 + 2 \alpha) \bar{l},
\nonumber \\
  |c_{2}|^{2}&=&\alpha \bar{l},
\label{enbuvv}
\end{eqnarray}
where $\alpha$ is a variational parameter. Minimizing the
interaction energy in Eq.\,(\ref{enbuv}) with respect to $\alpha$ we find that
$\alpha = 1$. More generally using similar arguments
we find that to leading order
\begin{eqnarray}
  |c_{m}|^{2} \propto \bar{l}^{|m-1|}, 
\label{l=1s}
\end{eqnarray}
and the explicit expressions for the coefficients are
\begin{eqnarray}
  |c_{0}|^{2}&=&2 \, \bar{l} - \frac 3 2 \, \bar{l}^{2},
  \nonumber \\
  |c_{1}|^{2}&=&1 - 3 \, \bar{l} + \frac {27} 8 \, \bar{l}^{2},
  \nonumber \\
  |c_{2}|^{2}&=&\bar{l} - \frac 9 4 \, \bar{l}^{2},
  \nonumber \\
  |c_{3}|^{2}&=&\frac 3 8 \, \bar{l}^{2},
  \nonumber \\
 {\rm and \phantom{XX}} |c_{4}|^{2}&=& \frac {{\bar l}^3} {12},
\label{cisl=1}
\end{eqnarray}
and the interaction energy to order ${\bar l}^{3}$ is,
\begin{eqnarray}
   \langle V \rangle = \left[ 
  \frac 1 4 + \frac {\bar l} 4 + {\cal O} \, ({\bar l}^4) \right]
 N^2 v_{0} + {\cal O}(N v_{0}) .
\label{intel1}
\end{eqnarray}
The above equation implies that also in the region $l < 1$
the interaction energy varies linearly with the angular momentum
to the order we have examined, which is also in agreement with
the numerical simulations.  The coefficient of the linear term is
the same as the one we found for small values of the angular momentum.
  
  Equations (\ref{intel00}) and (\ref{intel1}) as well as the numerical
results [see Fig.\,2] strongly suggest that 
the interaction energy $\langle V \rangle$ drops linearly as
a function of $L/N$ in the whole region $0 \le L/N \le 1$. 
The same result was derived by Butts and Rokhsar \cite{Rokhsar}
numerically within the mean-field approximation. 
In Ref.\,\cite{Bertsch} Bertsch and Papenbrock performed
exact diagonalizations of degenerate states of a given $L$ and
found that up to machine accuracy the energy of the lowest state
for a given $L$ varies linearly with $L$ in the range $2 \le L \le N$, 
in agreement with our analytic expansions. This result is specific to
the contact form of the effective interaction, and is probably connected
to a hidden symmetry in our Hamiltonian $H$, as discussed by Pitaevskii 
and Rosch \cite{LP}, where the same Hamiltonian was  
considered in the context of breathing modes.
\vskip1pc
{\bf b. Analytical results for $l > 1$}
\vskip1pc
   We turn now to the case $l > 1$. We calculate the difference
in energy between the states $|1^{N}\rangle$, and $|1^{N-1}, 
(\lambda+1)^{1} \rangle$, by a method similar
to that which for small $l$ led to Eqs.\,(\ref{states}) and (\ref{exce}).
The energy $\epsilon_{\lambda}$ of this 
$2^{\lambda}$-pole excitation with $L=N+\lambda$, with $\lambda \ll N$
is given by 
\begin{eqnarray}
    \epsilon_{\lambda} =  \lambda \hbar \omega
   - \frac 1 2 \left( 1 - \frac {\lambda+2} {2^{\lambda}} \right) N v_{0}
 + {\cal O} (v_0).
\label{exceu}
\end{eqnarray}
This formula implies that for $l>1$, the single-particle 
excitations with the lowest energy per unit of angular
momentum are those with $\lambda = 4$ or 5,
which means that the actual angular momentum carried by the particles is
$m = 5$ or 6. In contrast to the low-angular momentum case, here 
both $|c_5|^2$ and $|c_6|^2$ vary linearly with $\bar{l}$, 
$|c_5|^2 \propto |c_6|^2 \propto \bar{l}$, where $\bar{l} = L/N -1$.
In addition, in this regime we find that the energy has corrections 
of higher order than linear. Using similar arguments to those given
before for small $l$, we find, to order $\bar{l}^{2}$,
\begin{eqnarray}
    |c_{0}|^{2}&=& 0.1213 \, \bar{l}^{2},
\nonumber \\
  |c_{1}|^{2}&=&1 -  0.2241 \, \bar{l},
  \nonumber \\
   |c_{2}|^{2}&=& 0.1934 \, \bar{l}^{2},
\nonumber \\
   |c_{5}|^{2}&=& 0.1205 \, \bar{l},
\nonumber \\
   |c_{6}|^{2}&=& 0.1036 \, \bar{l},
\nonumber \\
   |c_{9}|^{2}&=& 1.7 \times 10^{-3} \, \bar{l}^2,
\nonumber \\
   |c_{10}|^{2}&=& 1.3 \times 10^{-3} \, \bar{l}^{2},
\nonumber \\
   |c_{11}|^{2}&=& 8.6 \times 10^{-3} \, \bar{l}^2,
\label{cisl=1p}
\end{eqnarray}
and the interaction energy is
\begin{equation}
      \langle V \rangle = \left[ \frac 1 4 - \frac {5 \, \bar l} {64}
     + 6.7 \times 10^{-3} \, \bar{l}^{2} + {\cal O}({\bar l}^3) \right]
      N^2 v_0  + {\cal O} (N v_0).
\label{intel1p}
\end{equation}
If we compare the expressions (\ref{intel1}) and (\ref{intel1p})
for the interaction energy we see that there is a change in its slope,
$\partial \langle V \rangle / \partial L$, as $L$ passes $N$, from 
$-N v_0/4$ for $L<N$ to $-5N v_0/64$ for $L>N$.

\subsection{Results for higher values of $L/N$}

    We mentioned earlier that as the angular momentum per particle 
increases even further, there are certain ranges of values of $L/N$ 
over which the state with the lowest energy has a specific symmetry.
The lowest value of $L/N$ for which this occurs is $\approx 1.75$
and the symmetry of the state is two-fold, i.e., only $c_{2m} \neq 0$. 
We have examined analytically as an example the case $L/N=2$.
Keeping only the first three non-zero coefficients, which are the 
dominant ones, we find to order $\bar{l} = L/N -2$ that
\begin{eqnarray}
  |c_{0}|^{2}&=&\beta(\bar{l}) - \bar{l}/4,
  \nonumber \\
  |c_{2}|^{2}&=&1-2 \beta(\bar{l}),
  \nonumber \\
 {\rm and \phantom{XX}} |c_{4}|^{2}&=&\beta(\bar{l}) + \bar{l}/4,
\label{cisl2}
\end{eqnarray}
where 
\begin{eqnarray}
   \beta(\bar{l})&=& \frac {3092 -48 \sqrt 6}{12695} + 
  \frac {17 (64 \sqrt 6 + 109)} 
 {10156} \bar{l} \nonumber \\
&\approx& 0.2343 + 0.4449 \bar{l},  
\label{expansionl=2n}
\end{eqnarray}
or
\begin{eqnarray}
  |c_{0}|^{2}&\approx&0.2343 + 0.1949 \bar{l},
  \nonumber \\
  |c_{2}|^{2}&\approx&0.5314 - 0.8897 \bar{l},
  \nonumber \\
  |c_{4}|^{2}&\approx&0.2343 + 0.6949 \bar{l},
\label{cisl2new}
\end{eqnarray}
and the corresponding interaction energy is:
\begin{eqnarray}
  \langle V \rangle = \left[ A - B \bar{l} + C \bar{l}^{2}
  + {\cal O} \, (\bar{l}^{3}) \right]
   N^2 v_0  
+ {\cal O} (N v_0),
\label{intel2}
\end{eqnarray}
where 
\begin{eqnarray}
   A &=& \frac 3 {16} + \frac {\beta_0} {32} (7 - 4 \sqrt 6)
 + \frac {\beta_0^2} {256} (64 \sqrt 6 - 109),
\nonumber \\
  B &=& \frac 1 {512} (4 + 85 \beta_0),
\label{intel5}
\end{eqnarray}
and $\beta_0 = \beta(0)$. The actual numbers which appear in 
Eq.\,(\ref{intel2}) are
\begin{eqnarray}
  \langle V \rangle \approx \left[ 0.1773 - 0.0467\bar{l}+0.0170\bar{l}^{2}
  + {\cal O} \, (\bar{l}^{3}) \right]
   N^2 v_0
 \nonumber \\ + {\cal O} (N v_0).
\label{intel6}
\end{eqnarray}
There is no change in the slope of the interaction energy as $L$ passes
$2N$. Figure 5 shows lines of constant density,
$|\Psi|^2=\rm{constant}$, for $L/N=2$. The occurence of two nodes
in the density reflects the presence of two displaced vortices,
and thus we see that the lowest-energy state of the system has 
two separated vortices and not a doubly quantized vortex. 
This clearly demonstrates the instability of the double-quantized
vortex state to formation of two vortices plus surface waves.

\section{Beyond the mean-field approximation}

   Another way of approaching the problem of rotation, is to diagonalize
the Hamiltonian within the space of degenerate states. This approach
goes beyond the mean-field 
approximation, since in mean field theory the many-body 
wavefunction is the product of the single-particle states,
whereas the diagonalization allows for the many-body state to have all
kinds of correlations between the particles. This technique can be
used by taking into account the whole set of states \cite{Bertsch}, but 
it is convenient and pedagogical to work in a restricted space, 
appropriately chosen. 

   As an example we consider small negative $L/N-1$. From the
analysis of Sec.\,III C we know that in this limit the states with 
$m = 1, 0$, and 2 are dominant. Therefore the eigenstates
\begin{eqnarray}
  |\mu,\tilde l \rangle = 
 |0^{\tilde l+\mu}, 1^{N-\tilde l-2\mu}, 2^{\mu} \rangle
\label{mu}
\end{eqnarray}
with $N$ particles and $L=N-\tilde l$ units of angular momentum are expected
to provide a good basis for describing the low-lying states for 
$\tilde l \ll N$. We restrict ourselves to this limit and demonstrate 
how one can derive an effective Hamiltonian which can be diagonalized
exactly. In the limit we
consider, $\tilde l$ is $\ll N$, and thus $\mu \sim \tilde l \ll N$.
The diagonal matrix elements in the Hamiltonian are, up to terms
of order $N$,
\begin{eqnarray}
  \langle \mu | V | \mu \rangle = \left( \frac 1 4 N (N-1)
 + \frac 1 2 \tilde l N + \frac 3 4 \mu N \right) v_0,
\label{diagb}
\end{eqnarray}
and the off-diagonal matrix elements are
\begin{eqnarray}
    \langle \mu+1 | V | \mu \rangle \approx
     \frac {\sqrt 2} 4 N v_0 \sqrt{(\mu + {\tilde l} +1) (\mu+1)}.
\label{meappr}
\end{eqnarray}
Ignoring for the moment the (diagonal) first term of Eq.\,(\ref{diagb}), 
which corresponds to the interaction energy of the state $|1^N \rangle$,
we see from Eqs.\,(\ref{diagb}) and (\ref{meappr}) that we have to 
diagonalize the Hamiltonian
\begin{eqnarray}
    {\tilde H} =  \left[ \frac 1 2 a_0^\dagger a_0 + \frac 1 4 a_2^\dagger a_2
    + \frac {\sqrt 2} 4 (a_2^\dagger a_0^\dagger + a_2 a_0) \right] N v_0,
\label{effh}
\end{eqnarray}
which can be done exactly by use of a Bogoliubov transformation.
Here $a_m$ is an annihilation operator that destroys a particle
with angular momentum $m \hbar$. Introducing the operators
$c$ and $d$ given by
\begin{eqnarray}
  c=a_0^\dagger + \sqrt 2 \, a_2; \, d = \sqrt 2 \, a_0 + a_2^\dagger, 
\label{bogdef}
\end{eqnarray}
we may write the Hamiltonian as
\begin{eqnarray}
   {\tilde H} =  \frac {N} {4} (d^\dagger d - 1) v_0.
\label{effhh}
\end{eqnarray}
Acting on states of the type (\ref{mu}) the operator 
$d^\dagger d - c^\dagger c$ is diagonal, and has an eigenvalue $N-L$.
Therefore $d^\dagger d$ can be eliminated and from Eq.\,(\ref{effhh}) 
we obtain
\begin{eqnarray}
  {\tilde H} =  \frac {N} {4} (N-L-1) v_0 + \frac {N v_0} 4 c^\dagger c.
\label{FinBog}
\end{eqnarray}
The total interaction energy is thus the eigenenergy of $\tilde H$ plus the
diagonal part $N (N-1) v_0/4$, or
\begin{eqnarray}
   \langle V \rangle &=& \frac N 4 (N-1) v_0 + \frac N 4 
  (N - L -1 + \langle c^\dagger c \rangle) v_0 + {\cal O} (v_0)
\nonumber \\
    &=& \frac {N (2N - L -2)} 4 v_0 + \frac 1 4 
  \langle c^\dagger c \rangle N v_0  + {\cal O} (v_0).
\label{limene}
\end{eqnarray}
In the ground state  $\langle c^\dagger c \rangle = 0$, and the energy
given by Eq.\,(\ref{limene}) 
is the same as that derived numerically in Ref.\,\cite{Bertsch}.  
The presence of the term $\langle c^\dagger c \rangle$ in Eq.\,(\ref{limene})
implies that the excited states in this limit of small negative $L/N-1$
are separated from the ground state by an amount 
\begin{eqnarray}
   \Delta E = \frac N 4  v_0 + {\cal O} (v_0).
\label{de}
\end{eqnarray}
 
We have also performed numerical diagonalization, and we have  
confirmed the above result (\ref{limene}), as well as
Eq.\,(\ref{de}).
The average occupancy of the states of the non-interacting problem is, 
for the lowest-energy state and $L=N$:
\begin{equation}
   |c_1|^2 = 1-\frac 2 N + {\cal O} \left( \frac 1 {N^2} \right); 
  \,  |c_0|^2=|c_2|^2 = \frac 1 N + {\cal O} \left( \frac 1 {N^2} 
 \right),
\label{occupdiag}
\end{equation}
and thus in the limit $N \rightarrow \infty$ there is agreement
between the mean-field approximation and the present one.

\section{Summary and conclusions}

   To summarize, we have studied the lowest-energy states of a system 
of rotating, weakly interacting harmonically trapped bosons. Within the
mean-field approximation, for $L/N \rightarrow 0$ we find that
the angular momentum is carried mainly by quadrupole $(|m| =2)$ 
excitations. We have demonstrated that diagrammatic
perturbation theory also leads to the same results as the method
we have used here.

  For $L/N = 1$ the angular momentum is carried by particles in the
$m=1$ state, while for small negative $L/N - 1$ the $m=0$ and $m=2$ states
are also populated. In the limits $L/N \rightarrow 0$ and $L/N \rightarrow
1$ the energy is a linear 
function of the angular momentum up to the order we have explored, while
numerically this linearity persists in the whole region $0 
\le L/N \le 1$. This result is specific to the contact form of the effective
interaction, and does not hold for more general interactions.

   For small positive $L/N-1$, the states which carry the additional 
angular momentum are those with $m=5$ and $m=6$. In addition,
as $L$ passes $N$ the derivative of the interaction energy with
respect to the angular momentum changes abruptly.
We have also found that for $L/N \approx 1.75$ there is a second order
phase transition and for $1.75 \le L/N \le 2.03$ the lowest energy
state has two-fold symmetry. At $L/N \approx 2.03$, there is a
first order phase transition to a state with
three-fold symmetry. More generally, for higher values of $L/N$
a vortex array develops.

   The Gross-Pitaevskii wavefunction is a power series in $\tilde z=x+iy$.
Thus if one truncates the series at $m=m_{\rm max}$, the wavefunction
will have $m_{\rm max}$ nodes. In the vicinity of a node at $\tilde z= 
\tilde z_0$ the wavefunction varies as $\tilde z- \tilde z_0$ and
therefore each node corresponds to a singly quantized vortex having 
the same sense as the total angular momentum. 
It is instructive to study how the vortex lines 
move as the angular momentum is increased. For low angular momentum, the
condensate wavefunction has only $m=0$ and $m=2$ components, and it
is therefore proportional to  $[1 - (l^{1/2}/2) (\rho/a_{\rm osc})^2 e^{2i\phi}]
\exp(-\rho^2/2a_{\rm osc}^2)$ for the choice of $\phi_2=\pi$, according
to Eq.\,(\ref{phases}) with $\phi_1 =0$. 
This has vortices on the $x$ axis at $x=\pm (2/l^{1/2})^{1/2} a_{\rm osc}$.
With increasing angular momentum, components of the
wavefunction with odd $m$ grow, and the two-fold symmetry of 
the cloud is broken, as may be seen in Fig.\,1 for $L/N=0.1$, 
one of the vortices moving to larger distances, and the other
to smaller ones. The $c_3$ term leads to a third vortex at large
distances from the origin. For $L/N=1$ there is only one vortex,
which is at the origin. With further increase in $L/N$, the
velocity field is at first still dominated by a vortex close to the origin,
but subsequently a second vortex moves into the cloud until at 
$L/N \approx 1.75$ the two-fold symmetry is restored. As $L/N$
increases towards the value 2.03, at which the first-order transition
to the state with three-fold symmetry mentioned above occurs, the
separation of the two vortices changes little. 

   In this paper we have also investigated effects not included in 
mean-field theory by diagonalizing a model Hamiltonian for $L$
close to, but less from, $N$. We find that for $L=N$ 
the occupancy of the $m=1$ state is 1, with corrections of order $1/N$.
We have calculated the energy up to terms of order $N$.
Finally we also found that the low-lying excited states
are separated from the lowest state by energies of order $N v_0$
of the same angular momentum.
  
    In this study we have examined the limit of weak interactions.  When
the interaction energy per particle $nU_0$ becomes comparable to or
greater than $\hbar \omega$, components of the wavefunction that are
not members of the lowest multiplet in the absence of interaction
must be included.  Calculations for this regime based on the 
Gross-Pitaevskii equation have been carried out by Isoshima and Machida
\cite{IM}. Comparison of our results with theirs is difficult
because these authors calculated the lowest energy state in a rotating frame,
rather than the lowest energy state for a given angular momentum.

    One question of importance both conceptually and because of its
relevance to experiment is whether or not the states are stable to
small perturbations, and if they are not, what is the lifetime of the
state.  The answer to these questions depends on the nature of the
perturbation, whether it is due to a deformation of the trap, or
to interactions with particles outside the condensate, and we
shall discuss it elsewhere.

    In our calculations above we have shown for a particular example that
the Gross-Pitaevskii approach gives correctly the contribution to the
energy of order $N^2$.  This result, which is alluded 
to in Ref.\,\cite{Rokhsar}, is more general, and in a future
publication \cite{JKMR} it will be shown how the Gross-Pitaevskii approach is
recovered as the first term in an expansion in powers of $1/N$.  The
method may be extended to calculate contributions to the energy of
order $N$ which are in excellent agreement with results obtained by
numerical diagonalization of the Hamiltonian.

\begin{appendix}

\section*{Perturbation theory approach}

We show in this appendix that one can use perturbation theory to
derive an effective Hamiltonian in the region $L/N \rightarrow 0$,
corresponding to Eq.\,(\ref{lsqreff}). We assume that only the
states with $m=0, 2$, and 3 are macroscopically occupied. However,
other states (the $m=1, 4, 5$ and 6 in this case)
give corrections to the energy that can be 
treated perturbatively. Let us demonstrate how this works
by considering the interaction energy up to $l^2$. As long as 
both $c_2$ and $c_3$ vary as $l^{1/2}$, the only processes that 
contribute to the interaction energy up to $l^2$ are shown in Fig.\,7.

Let us consider the first process on the left as an example.
The matrix element $M$ corresponding to the vertex, where two particles
with $m=2$ scatter to states with $m=0$ and $m=4$,
is equal to 
\begin{eqnarray}
  M = \frac {\sqrt{6}} {16} N_2 \sqrt N v_0. 
\label{me}
\end{eqnarray}
From Eq.\,(\ref{exce})
we find that the difference in the energy between the intermediate
state and the initial state is 
\begin{eqnarray}
  \delta \epsilon = - \frac {N v_0} {8} + {\cal O} (v_0), 
\label{deltae}
\end{eqnarray}
and thus perturbation theory implies that the correction to the energy is
\begin{eqnarray}
  \frac {|M|^2} {\delta \epsilon} = - \frac 3 {16} N_2^2 v_0
 = - \frac 3 {16} |c_2|^4 N^2 v_0,
\label{pert}
\end{eqnarray}
which is precisely the correction $\Delta {\cal E}$
given by Eq.\,(\ref{corrc4}) (plus terms of order $N v_0$).
Similarly the other diagrams shown below give $-5 |c_3|^4/544$,
$- |c_2|^2 |c_3|^3/8$, and $- 3|c_2|^2 |c_3|^2/4$ in units of $N^2 v_0$,
respectively, and are identical to the corrections given by the terms in
the last four lines of Eq.\,(\ref{lsqr}).
\begin{figure}
\begin{center}
\epsfig{file=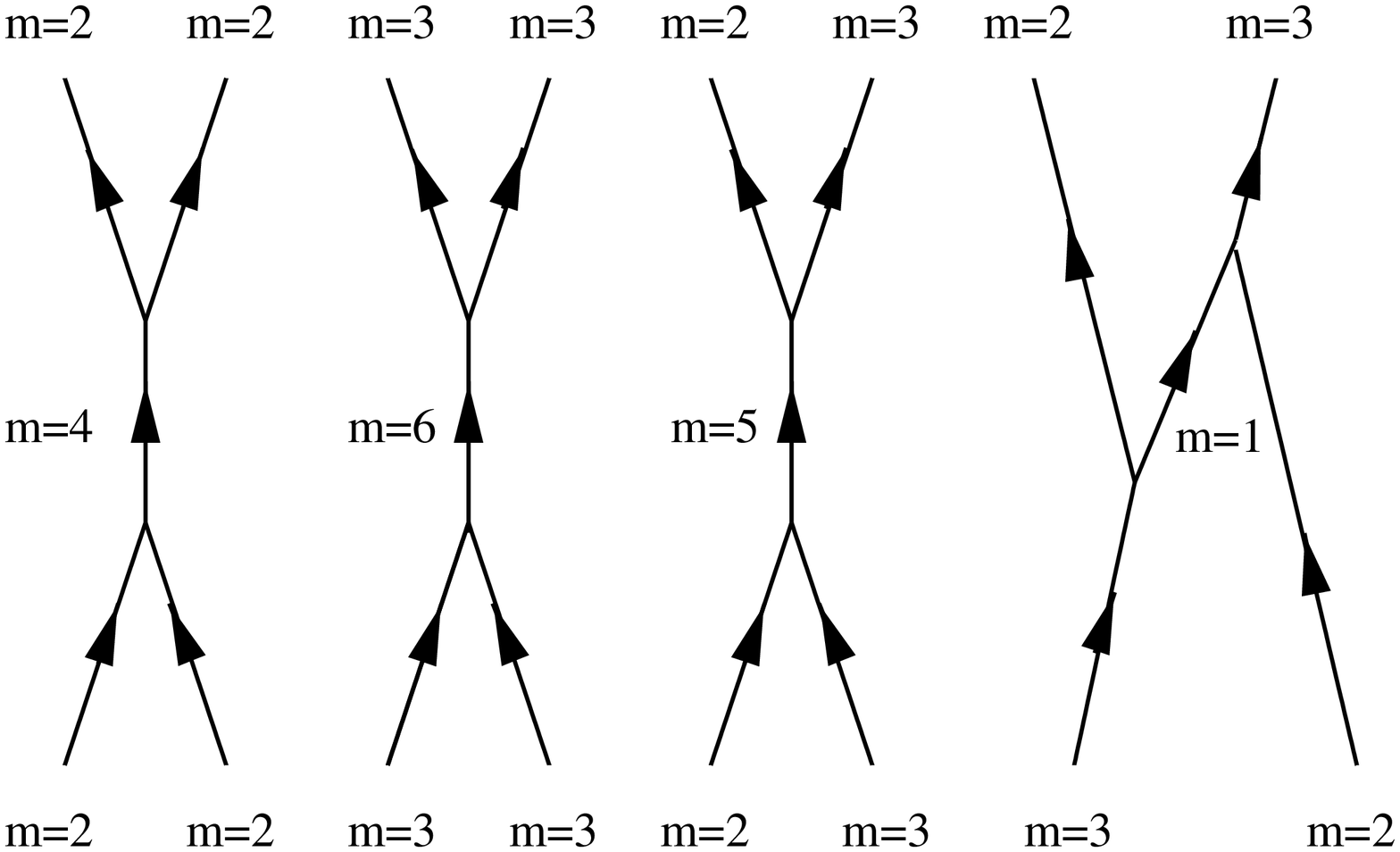,width=8.0cm,height=4.0cm,angle=0}
\vskip1pc
\begin{caption}
{The four diagrams contributing to the interaction energy 
to order $l^2$.}
\end{caption}
\end{center}
\label{FIG5}
\end{figure}

\end{appendix}
 
\vskip0.5pc

   G.M.K. was supported by the European Commission, TMR program,
contract No.\,ERBFMBICT 983142. Helpful discussions with A. Jackson
and S. Reimann are gratefully acknowledged. G.M.K. would like
to thank the Foundation of Research and Technology, Hellas
(FORTH) for its hospitality.

\end{document}